\documentstyle[10pt]{article}

\newfont{\tenof}{msym10}

\newcommand{\F}{\Phi_{\sigma}}
\newcommand{\f}[1]{\varphi^{[f]}_{#1}}
\newcommand{\V}[2]{V^{[f]}_{#1}(#2)}
\newcommand{\ts}[1]{\tau^{(#1)}_{\sigma}}
\newcommand{\ls}[1]{\lambda_{\sigma(#1)}}
\newcommand{\sumij}{\sum_{\scriptstyle i,j \atop \scriptstyle i\ne j}}
\newcommand{\csch}{\mathop{\rm csch}\nolimits}

\begin{document}
%
%
\begin{center}
{\bf GENERALIZATION OF A RESULT OF MATSUO AND CHEREDNIK TO THE
CALOGERO-SUTHERLAND-MOSER INTEGRABLE MODELS WITH EXCHANGE
TERMS\footnote{Presented at the 4th Colloquium ``Quantum Groups and Integrable
Systems'', Prague, 22--24 June 1995}}

\bigskip\bigskip
{CHRISTIANE QUESNE\footnote{Directeur de recherches
FNRS}}

\bigskip
{\sl Physique Nucl\'eaire Th\'eorique et Physique Math\'ematique, Universit\'e
Libre
de Bruxelles, Campus de la Plaine CP229, Boulevard du Triomphe, B-1050
Brussels, Belgium}

\bigskip
{\bf Abstract}
\end{center}

A few years ago, Matsuo and Cherednik proved that from some solutions of the
Knizhnik-Zamolodchikov (KZ) equations, which first appeared in conformal field
theory, one can obtain wave functions for the Calogero integrable system. In
the
present communication, it is shown that from some solutions of generalized KZ
equations, one can construct wave functions, characterized by any given
permutational symmetry, for some Calogero-Sutherland-Moser integrable models
with exchange terms. Such models include the spin generalizations of the
original
Calogero and Sutherland ones, as well as that with $\delta$-function
interaction.
\par
%
%
\section{Introduction}
The Calogero integrable system~\cite{cal1} consists of $N$ nonrelativistic
particles
on the line interacting through a two-body potential of the inverse square
type,
\begin{equation}
  V_C^{(k)} = k(k-1) \sum_{\scriptstyle i,j=1 \atop \scriptstyle i\ne j}^N
  \frac{1}{(x_i-x_j)^2}.             \label{eq:1calpot}
\end{equation}
Together with its trigonometric, hyperbolic, and elliptic
generalizations~\cite{cal2}--\cite{mos}, it constitutes the
Calogero-Sutherland-Moser (CSM) integrable model, which is related to the root
systems of ${\cal A}_{N-1}$ algebras~\cite{pere} and is founding some
interesting
applications in field-theoretical and condensed-matter contexts.\par
%
%
A breakthrough in the study of  CSM systems occurred some three years ago
when Polychronakos~\cite{poly} and Brink {\sl et al}~\cite{brink1}
independently introduced an exchange operator formalism, leading to a set of
$N$
commuting first-order differential operators, known in the  mathematical
literature
as Dunkl operators~\cite{dunkl}. Such operators led to Hamiltonians with
exchange
terms, connected with the spin generalizations of the CSM systems~\cite{ber}.
Spectra and wave functions of the latter can be obtained by simultaneously
diagonalizing the $N$ commuting Dunkl operators.\par
%
%
At approximately the same time, another approach to CSM systems made its
appearance. It was based upon the Knizhnik-Zamolodchikov (KZ) equations,
previously
introduced in conformal field theory~\cite{kz}. The latter can be written as
\begin{equation}
  \partial_i\Phi = \left(k \sum_{j\ne i} \frac{P^{(ij)}}{x_i-x_j} +
\lambda^{(i)}
  \right) \Phi, \qquad i=1,2,\ldots,N,               \label{eq:kz}
\end{equation}
where $\Phi = \Phi(x_1,x_2,\ldots,x_N)$ takes values in the tensor product
$V\otimes V\otimes\cdots\otimes V = V^{\otimes N}$ of some $N$-dimensional
vector space $V$, $P^{(ij)}$ is the permutation of the $i^{\rm th}$ and $j^{\rm
th}$
factors, $\lambda = \hbox{\rm diag}(\lambda_1, \ldots, \lambda_N)$ is a
diagonal
matrix considered as a parameter, $\lambda^{(i)}$ is the operator in
$V^{\otimes N}$
acting as $\lambda$ on the $i^{\rm th}$ factor and identically on all other
factors.\par
%
%
Matsuo~\cite{mat} and Cherednik~\cite{che} indeed showed that if one considers
solutions of~(\ref{eq:kz}) that can be written as
\begin{equation}
  \Phi = \sum_{\sigma\in S_N} \F e_{\sigma}, \qquad e_{\sigma}
    = e_{\sigma(1)} \otimes e_{\sigma(2)} \otimes \cdots \otimes
    e_{\sigma(N)}, \label{eq:Phi}
\end{equation}
where $S_N$ is the symmetric group, and $e_k$ denotes a column vector with
entry~1 in row~$k$ and zeroes everywhere else, then the symmetric and
antisymmetric combinations
\begin{equation}
  \varphi^{[N]} = \sum_{\sigma \in S_N} \Phi_{\sigma}, \qquad \varphi^{[1^N]} =
  \sum_{\sigma \in S_N} (-1)^{\sigma} \Phi_{\sigma},   \label{eq:sa}
\end{equation}
of the components~$\Phi_{\sigma}$ of function~(\ref{eq:Phi}), where
$(-1)^{\sigma}$ denotes the parity of permutation~$\sigma$, are eigenfunctions
with
eigenvalue $- \sum_i \lambda_i^2$ of the Calogero Hamiltonian,
\begin{equation}
  \left(- \Delta + V_C^{(k)} + \sum_i \lambda_i^2\right) \varphi^{[N]} =
  \left(- \Delta + V_C^{(k+1)} + \sum_i \lambda_i^2\right) \varphi^{[1^N]} = 0,
  \label{eq:hcal}
\end{equation}
corresponding to parameters~$k$ and~$k+1$, respectively.
\par
%
%
In the present communication, we report on some new results~\cite{cq} extending
those of Matsuo and Cherednik to some CSM systems with exchange terms, thereby
providing new links between the latter and some generalized KZ~equations. Such
equations are reviewed in Sec.~2, and used in Sec.~3 to construct wave
functions for
CSM systems with exchange terms. Section~4 contains the conclusion.\par
%
%
\section{Generalized Knizhnik-Zamolodchikov equations}
Let us generalize system~(\ref{eq:kz}) into
\begin{equation}
  \partial_i\Phi = \left(\sum_{j\ne i} \left(f_{ij}(x_i-x_j) P^{(ij)} +
    c\, T^{(ij)}\right) + \lambda^{(i)}\right) \Phi, \qquad i=1,2,\ldots,N,
    \label{eq:genkz}
\end{equation}
where all symbols keep the same meaning as in~(\ref{eq:kz}), but $k
(x_i-x_j)^{-1}$
is replaced by a yet undetermined function $f_{ij}(x_i-x_j)$, and there is an
additional term proportional to an operator $T^{(ij)}$ acting only on the
$i^{\rm th}$
and $j^{\rm th}$ factors, and such that $T$ is the following operator on
$V\otimes
V$:
\begin{equation}
  T = \sum_{k>l} \left(E_{kl}\otimes E_{lk} - E_{lk}\otimes E_{kl}\right).
    \label{eq:T}
\end{equation}
Here $E_{kl}$ denotes the $N\times N$ matrix with entry~1 in row~$k$ and
column~$l$ and zeroes everywhere else (for a previous use of the
operator~$T^{(ij)}$, see e.g.~\cite{fel} and references therein).\par
%
%
Let us again restrict ourselves to solutions of~(\ref{eq:genkz}) that can be
written
in the form given by~(\ref{eq:Phi}). For such functions, Eq.~(\ref{eq:genkz})
is
equivalent to the set of equations
\begin{equation}
  \partial_i\F = \sum_{j\ne i} \left(f_{ij}(x_i-x_j) + c\, \ts{ij}\right)
  \Phi_{\sigma \circ p_{ij}} + \ls{i} \F, \qquad i=1,2,\ldots,N,
\label{eq:genkzcom}
\end{equation}
where $\sigma$ is an arbitrary permutation of $S_N$, $p_{ij}$ denotes the
transposition of~$i$ and~$j$, and $\ts{ij} \equiv \hbox{\rm sgn}\bigl(\sigma(i)
-
\sigma(j)\bigr)$ satisfies the relations
\begin{eqnarray}
  & & \ts{ij} = - \tau^{(ij)}_{\sigma \circ p_{ij}} = - \ts{ji}, \qquad
      \tau^{(ik)}_{\sigma \circ p_{ij}} = \ts{jk}, \qquad
      \tau^{(kl)}_{\sigma \circ p_{ij}} = \ts{kl}, \nonumber \\
  & & \ts{ij} \ts{ik} + \ts{jk} \ts{ji} + \ts{ki} \ts{kj} = 1,
      \label{eq:tau}
\end{eqnarray}
for any $i\ne j \ne k \ne l$. In deriving Eq.~(\ref{eq:genkzcom}), use has been
made of
the properties that the operators $P^{(ij)}$, $T^{(ij)}$, and~$\lambda^{(i)}$
transform
the components $\F$ into $\Phi_{\sigma \circ p_{ij}}$, $\ts{ij} \Phi_{\sigma
\circ
p_{ij}}$, and $\ls{i} \F$, respectively.\par
%
%
It can be easily shown~\cite{cq} that the integrability conditions
of~(\ref{eq:genkzcom}), i.e., $\partial_j \partial_i \Phi_{\sigma} = \partial_i
\partial_j \Phi_{\sigma}$ for any $i$, $j=1$,~2, $\ldots$,~$N$, and any $\sigma
\in
S_N$, are satisfied if and only if
\begin{equation}
  f_{ij}(u) = - f_{ji}(-u), \label{eq:ic1}
\end{equation}
for any $i$, $j$, such that $i<j$, and
\begin{equation}
  f_{ij}(u) f_{jk}(v) - f_{ik}(u+v) \left[f_{ij}(u) + f_{jk}(v)\right] =
  - c^2, \label{eq:ic2}
\end{equation}
for any $i$, $j$,~$k$, such that $1\le i<j<k\le N$.\par
%
%
Equation~(\ref{eq:ic2}) looks like a functional equation first considered by
Sutherland~\cite{suth}, and solved by Calogero~\cite{cal3} through a small-$x$
expansion. A similar procedure can be used here to derive all the solutions
of~(\ref{eq:ic2}) that are odd and meromorphic in a neighbourhood of the
origin~\cite{cq}. Among the latter, one finds
\begin{equation}
  f_{ij}(u) = f_{ji}(u) = F(u), \qquad 1\le i<j\le N, \label{eq:mero}
\end{equation}
where $F(u)$ denotes the function
\begin{equation}
  F(u) = \cases{
      k \omega \coth \omega u & if $c^2 = k^2\omega^2 >0$, \cr
      \noalign{\smallskip}
      k/u                     & if $c^2 = 0$, \cr
      \noalign{\smallskip}
      k \omega \cot \omega u  & if $c^2 = -k^2\omega^2 <0$, \cr}
  \label{eq:F}
\end{equation}
and $N$ may take any value such that $N\ge3$. For $c^2=0$,
function~(\ref{eq:mero}) is the unique solution of Eq.~(\ref{eq:ic2}) that is
odd and
meromorphic in a neighbourhood of the origin, while for $c^2\ne0$, there is
another
solution, for which not all $f_{ij}(u)$'s are equal, and which will not play
any role in
the next Section.\par
%
%
Eq.~(\ref{eq:ic2}) also has some singular solutions, such as
\begin{equation}
  f_{ij}(u) = f_{ji}(u) = c\, \mbox{sgn}(u) = c\, [\theta(u) - \theta(-u)],
  \qquad 1\le i<j\le N, \label{eq:sing}
\end{equation}
where $\theta(u)$ denotes the Heaviside function.\par
%
%
\section{Generalization of the result of Matsuo and Che\-rednik}
{}From a set of $N!$ functions $\F(x_1,\ldots,x_N)$, $\sigma \in
S_N$, satisfying Eq.~(\ref{eq:genkzcom}), one can construct in general $N!$
functions $\f{rs}(x_1,\ldots,x_N)$, defined by
\begin{equation}
  \f{rs} = \sum_{\sigma\in S_N} \V{rs}{\sigma} \F, \label{eq:mixed}
\end{equation}
where $[f] \equiv \left[f_1 f_2 \ldots f_N\right]$ runs over all $N$-box
Young diagrams, $r$ and $s$ label the standard tableaux associated with
$[f]$, arranged in lexicographical order, and $\V{rs}{\sigma}$ denotes
Young's orthogonal matrix representation of $S_N$~\cite{ruth}. For $[f] = [N]$
or
$\bigl[1^N\bigr]$, since $V^{[N]}(\sigma) = 1$ and $V^{[1^N]}(\sigma) =
(-1)^{\sigma}$, functions~(\ref{eq:mixed}) reduce to those considered by Matsuo
and
Cherednik and given in~(\ref{eq:sa}).\par
%
%
By using the previous equations, as well as some elementary results about
transpositions,
\begin{equation}
  p_{ik} \circ p_{ij} = p_{ij} \circ p_{jk} = p_{jk} \circ p_{ik},
  \label{eq:trans}
\end{equation}
and matrix representations,
\begin{equation}
  \V{rs}{\sigma \circ \sigma'} = \sum_t \V{rt}{\sigma} \V{ts}{\sigma'},
  \qquad \V{rs}{1} = \delta_{r,s}, \label{eq:matrep}
\end{equation}
it is straightforward to show~\cite{cq} that the functions $\f{rs}$ satisfy the
system of equations
\begin{eqnarray}
  \partial_i \f{rs} & = & \sum_{j\ne i} f_{ij} \sum_t \f{rt} \V{ts}{p_{ij}}
     - c \sum_{j\ne i} \sum_t \biggl(\sum_{\sigma} \ts{ij} \V{rt}{\sigma} \F
     \biggr) \V{ts}{p_{ij}} \nonumber \\
  & & \mbox{} + \sum_{\sigma} \ls{i} \V{rs}{\sigma} \F, \qquad
     i=1,2,\ldots,N, \label{eq:first}
\end{eqnarray}
and that their Laplacian is given by
\begin{equation}
  \Delta \f{rs} = \left(\sumij \left(f_{ij}^2(x_i-x_j) + \left(\partial_i
  f_{ij}(x_i-x_j)\right) K_{ij} - c^2\right) + \sum_i \lambda_i^2\right)
  \f{rs}. \label{eq:laplacian}
\end{equation}
In~(\ref{eq:laplacian}), $K_{ij} = K_{ji}$, $1\le i<j\le N$, are some
operators, whose
action on $\f{rs}$ is defined by
\begin{equation}
  K_{ij} \f{rs} = \sum_t \f{rt} \V{ts}{p_{ij}}. \label{eq:K}
\end{equation}
\par
%
%
In the special cases where $[f] = [N]$ or $\bigl[1^N\bigr]$, the operators
$K_{ij}$
behave as $I$ or $-I$, respectively. Hence, for $f_{ij}$ given
by~(\ref{eq:mero}) and~(\ref{eq:F}), where $c^2=0$, Eq.~(\ref{eq:laplacian})
reduces
to Eq.~(\ref{eq:hcal}), i.e., Matsuo and Cherednik's result. Let us emphasize
that
Eq.~(\ref{eq:laplacian}) is also valid in the mixed-symmetry cases, and for any
function $\f{rs}$ constructed from any solution of~(\ref{eq:genkzcom}) via
transformation~(\ref{eq:mixed}).\par
%
%
As for $[f] \ne [N]$, $\bigl[1^N\bigr]$, the operators $K_{ij}$ have a rather
complicated effect on the functions $\f{rs}$, it is convenient to restrict the
latter
so that for any $i<j$, $K_{ij}$ may be interpreted as a permutation operator
acting on
the variables $x_i$ and $x_j$, and leaving the remaining variables $x_k$
unchanged,
\begin{equation}
  K_{ij} x_j = x_i K_{ij}, \qquad K_{ij} x_k = x_k K_{ij} \qquad k\ne i,j.
  \label{eq:xpermute}
\end{equation}
{}From~(\ref{eq:K}), it results that the conditions to be fulfilled by $\f{rs}$
are
\begin{eqnarray}
  \lefteqn{\f{rs}(x_1,\ldots,x_j,\ldots,x_i,\ldots,x_N)} \nonumber \\
  & = &\sum_t \f{rt}(x_1,\ldots,x_i,\ldots,x_j,\ldots,x_N) \V{ts}{p_{ij}},
      \qquad 1\le i<j\le N. \label{eq:condphi}
\end{eqnarray}
In terms of the components~$\F$ of~(\ref{eq:Phi}), such conditions amount to
\begin{eqnarray}
  \lefteqn{\F(x_1,\ldots,x_j,\ldots,x_i,\ldots,x_N)} \nonumber \\
  & = & \Phi_{\sigma\circ p_{ij}}(x_1,\ldots,x_i,\ldots,x_j,\ldots,x_N),
      \qquad 1\le i<j\le N, \label{eq:condPhi}
\end{eqnarray}
for any $\sigma\in S_N$. By differentiating both sides of~(\ref{eq:condPhi})
with
respect to~$x_k$ and using~(\ref{eq:genkzcom}) to calculate the derivatives,
one
finds that Eqs.~(\ref{eq:genkzcom}) and~(\ref{eq:condPhi}) are  compatible if
and only
if all functions $f_{ij}(u)$, $i\ne j$, coincide, hence in cases such
as~(\ref{eq:mero})
and~(\ref{eq:sing}).\par
%
%
For instance, when $f_{ij}$ is given by~(\ref{eq:mero}) and~(\ref{eq:F}), where
$c^2>0$ (hyperbolic case), Eq.~(\ref{eq:laplacian}) becomes
\begin{equation}
  \left(- \Delta + \omega^2 \sumij \left(\csch \omega(x_i-x_j)\right)^2
  k(k-K_{ij}) + \sum_i \lambda_i^2\right) \f{rs} = 0, \label{eq:suthex}
\end{equation}
with $K_{ij}$ defined by~(\ref{eq:xpermute}). Such an equation (as well as
similar
results for the remaining cases) shows that from any solution of type
(\ref{eq:Phi}),~(\ref{eq:condPhi}) of the KZ equations~(\ref{eq:genkz}), with
$f_{ij}$
given in~(\ref{eq:mero}), one can obtain eigenfunctions $\f{rs}$ of the CSM
Hamiltonians~\cite{cal1}--\cite{mos} with exchange terms~\cite{ber}, which
are characterized by any given permutational symmetry $[f]$ under particle
coordinate exchange. To obtain wave functions describing an $N$-boson (resp.\
$N$-fermion) system, it only remains to combine $\f{rs}$ with a spin function
transforming under the same (resp.\ conjugate) irreducible
representation~$[f]$ (resp.~$[\tilde f]$) under exchange of the spin
variables. A similar result is valid for the Hamiltonian with
delta-function interactions~\cite{lieb}, corresponding to the functions
$f_{ij}$ given in~(\ref{eq:sing}).\par
%
%
\section{Conclusion}
By extending the Matsuo and Cherednik results to some integrable models with
exchange terms and wave functions of any permutational symmetry, we showed that
there exists a strong interplay between such models and (generalized)
KZ~equations.
As already noted by Brink and Vasiliev~\cite{brink2}, this also hints at some
deep
relationship between the latter and Dunkl operators.\par
%
%
Whether some results similar to those presented here also hold true for
elliptic CSM
models, and for integrable models related to root systems of algebras different
from
${\cal A}_{N-1}$~\cite{pere}, remains an interesting open question.
\par
%
%

\end{document}